\documentclass[superscriptaddress,twocolumn]{revtex4-2}
\usepackage[applemac]{inputenc}
\usepackage{amssymb}
\usepackage{natbib}
\usepackage{amsmath}
\usepackage{amsfonts}
\usepackage{graphicx}
\usepackage{subfigure}
\usepackage{mathrsfs}
\usepackage{dcolumn} 
\usepackage{bm}
\usepackage{color}
\usepackage{cancel}
\usepackage{mathbbol}
\usepackage{epsfig}
\usepackage{units}
\usepackage{esint}
\usepackage{soul} 
\usepackage{url}
\usepackage{afterpage}
\usepackage{hyperref}

\def\Tr{\mbox{Tr}}
\newcommand{\ket}[1]{\left\vert#1\right\rangle}
\newcommand{\bra}[1]{\left\langle#1\right\vert}
\newcommand{\av}[1]{\langle#1\rangle}

\begin{document}

\title{Experimental verification of fluctuation relations with a quantum computer} 
 
\author{Andrea Solfanelli}
\email{asolfane@sissa.it}
\affiliation{SISSA, via Bonomea 265, I-34136 Trieste, Italy}

\author{Alessandro Santini}
\email{asantini@sissa.it}
\affiliation{SISSA, via Bonomea 265, I-34136 Trieste, Italy}

\author{Michele Campisi}
\email{michele.campisi@nano.cnr.it}
\affiliation{NEST, Istituto Nanoscienze-CNR and Scuola Normale Superiore, I-56127 Pisa, Italy}
\address{Department of Physics and Astronomy, University of Florence, I-50019, Sesto Fiorentino (FI), Italy}
\affiliation{INFN - Sezione di Pisa, I-56127 Pisa, Italy}

\date{\today} 
 
\begin{abstract} 
Inspired by the idea that quantum computers can be useful in advancing basic science, we use a quantum processor to experimentally validate a number of theoretical results in non-equilibrium quantum thermodynamics, that were not (or were very little) corroborated so far. In order to do so, we first put forward a novel method to implement the so called two point measurement scheme, which is at the basis of the study of non-equilibrium energetic exchanges in quantum systems. 
Like the well-established interferometric method, our method uses an ancillary system, but at variance with it, it provides direct access to the energy exchange statistics, rather than its Fourier transform, thus being extremely more effective.
We first experimentally validate our ancilla-assisted two point measurement scheme, and then apply it to  i) experimentally verify that fluctuation theorems are robust against projective measurements, a theoretical prediction which was not validated so far,  ii) experimentally verify the so called heat engine fluctuation relation, by implementing a SWAP quantum heat engine. iii) experimentally verify that the heat engine fluctuation relation continues to hold in presence of intermediate measurements, by implementing the design at the basis of the so called quantum-measurement-cooling concept.
For both engines, we report the measured average heat and work exchanged and single out their operation mode. Our experiments constitute the experimental basis for the understanding of the non-equilibrium energetics of quantum computation and for the implementation of energy management devices on quantum processors.
\end{abstract}  
 
\keywords{quantum heat engines; quantum computation; quantum thermodynamics} 
 
\maketitle

\section{Introduction}

With the tremendous and fast advancements of quantum technologies, quantum computers have recently become a reality. Their development proceeds fast both in terms of the increasing number of quantum logical units (the qubits) that compose the processors, and in terms of the decreasing error accompanying the quantum information processing \cite{Alexeev21PRXQ2}. In the effort of improving the efficacy of quantum processors, understanding and mastering the disturbing thermal effects occurring during their operation is of crucial importance. In this regard quantum computation can largely  benefit from a field of investigation that has become recently known as quantum thermodynamics \cite{Goold16JPA49,Deffner19Book,Millen16NJP18}. Quantum thermodynamics is concerned with all the thermodynamic phenomena that may occur at the quantum level, ranging from thermal transport, to the fluctuations of thermodynamic quantities (notably heat and work), and in the way those can be mastered, e.g., by realising nano-scale quantum heat engines and refrigerators.

Most notably, not only can quantum computing benefit from quantum thermodynamics, but also the latter can benefit from the former:
A number of theoretical results in quantum thermodynamics, that can be experimentally checked with current quantum processors, are in fact still little if not at all corroborated by experimental evidence. Among them are the fluctuation relation for arbitrary open quantum systems \cite{Campisi09PRL102}, the robustness of fluctuation relations against perturbation induced by projective measurements \cite{Campisi10PRL105,Campisi11PRE83}, the fluctuation relation for heat engines \cite{Campisi14JPA47} \footnote{The very first experiment has been reported as we write, see Ref. \cite{Denzler21arXiv:2104.13427}.}, 
the realisation of two-qubit/two stroke engines, e.g., the SWAP engine \cite{Campisi15NJP17}, and the so called quantum measurement cooling \cite{Buffoni19PRL122}, whereby a refrigeration mechanisms is set by the very action of quantum measurement, without the aid of any feedback mechanism. 
With this work we substantially fill that gap by reporting the results of a number of new experiments that we performed on IBM quantum processors.  

All our experiments are based on the so-called two-point measurement scheme \cite{Campisi11RMP83} whereby the energy of the quantum system of interest is measured before and after an interaction with other agents (e.g., an external work source, another quantum system, a thermal environment or a measuring apparatus) has taken place. One problem that needs to be faced when implementing such a scheme is that often projective measurements are so invasive that they destroy the possibility to continue the experiment after the outcome of the measurement has been recorded. One way to circumvent this problem is to implement a Ramsey-like interferometric schemes where the information on the statistics of energy changes that would be obtained by subtracting final and initial measurement outcome are encoded in the state of an ancillary qubit \cite{Dorner13PRL110,Mazzola13PRL110} that is probed only at the end of the protocol. One drawback of this method is that it actually measures the characteristic function of the wanted statistics. That function needs to be sampled at a great number of points, in order for its inverse Fourier transform (i.e., the wanted statistics) to be efficiently extracted, and a great number of runs of the same experimental protocol must be repeated in order to collect sufficient statistics to achieve a good estimation of the characteristic function at each specific value of its argument.

Another problem that one typically faces when experimentally addressing quantum thermodynamics results is that normally they are based on the assumption that the quantum systems that are being manipulated and measured are initially in an equilibrium thermal state, while quantum technologies, including quantum processors, typically allow for the initialisation of quantum systems in a specific pure state. This problem is customarily circumvented by randomly initialising the system in a certain eigenstate of energy with the according Gibbs probability, thus emulating, rather than creating, a genuine thermal state. This strategy, that might be considered as not fully satisfactory has been implemented for example in \cite{An15NATPHYS11,Buffoni20QST5,Hernandez21NJP23}, and has the advantage of being immune from the initial measurement issue mentioned above \footnote{since  the system is  prepared in each run of the experiment in a specific energy-eigenstate, the initial energy measurement become superfluous}.

Here we propose a new method to effectively implement the two measurement scheme on a thermal state, that at once overcomes both issues in a most effective way. Like the interferometric method, our method, which we call the ancilla-assited two-point-measurement (AATPM) scheme, uses an ancilla qubit, but gives the energy change statistics directly (not its Fourier transform), thus requiring an enormously smaller amount of processor time to be efficiently executed, compared to the interferometric method. The ancilla, in our method serves as well to initialise the system in a thermal state, which at once cancels the necessity of emulating it. As will be detailed below, the trick is to prepare system and ancilla in a state that is the purification of the wanted thermal state for the system alone.

We implement our method on IBM quantum processors, and illustrate its employment for experimentally studying a number of problems in non-equilibrium quantum thermodynamics. We remark that IBM has recently added the possibility to insert projective measurement within a  quantum circuit (not only at its end), thus allowing for a direct implementation of the two- (in fact many-) point measurement scheme, a possibility that is not yet available on other existing quantum computing platforms, which we used to experimentally validate our method.

In Sec. \ref{sec:AATPM} we illustrate the theory at the basis of our AATPM scheme,
implement it on IBM quantum processors and validate its efficacy against the native IBM scheme, 
by using the accuracy with which the Jarzynski identity is reproduced as an experimental benchmark.
In Sec \ref{sec:projective_measurements} we experimentally verify the validity of the fluctuation relation in presence of intermediate projective measurements, a result that has been predicted in Ref. \cite{Campisi11PRE83,Campisi10PRL105}.
In Sec. \ref{sec:swap_quantum_heat_engine} we report the implementation of the two-stroke two-qubit SWAP engine \cite{Campisi15NJP17} and use it to verify the quantum heat engine fluctuation relation predicted in \cite{Campisi14JPA47}. 
We characterise the mode of operation of the device by measuring the actual energy exchanges occurring in the engine.
In Sec. \ref{sec:QMC} we report on the implementation of the heat engine design at the basis of the so-called quantum measurement cooling  \cite{Buffoni19PRL122}, whereby a qubit is refrigerated as a consequence of the act of being measured along with a hotter qubit (in an appropriate measurement basis). We experimentally validate the prediction that the heat engine fluctuation relation   should be obeyed in this case as well, and characterise the mode of operation of the device as we did for the SWAP engine design.

\section{Ancilla assisted two-point measurement: theory and experimental validation}\label{sec:AATPM}
\subsection{Theory}
We start by considering the prototypical non-equilibrium quantum thermodynamics scenario of a quantum system prepared in a thermal state at some inverse temperature $\beta$ and being subjected to a measurement-driving-measurement (MDM) protocol, where the measurement stages are projective measurements of the system Hamiltonian, and the driving stage is the application of some external forcing that induces some unitary evolution $U$ \cite{Campisi11RMP83}. The extension to more complex situations where, for example, the system is multipartite and each part is initialised at its own initial inverse temperature $\beta_i$, or the evolution is not unitary (e.g. a unitary interrupted by projective measurements) is straightforward.

We are interested in the statistics $\mathcal{P}(W)$ of the energy change of our quantum system (i.e., the work $W$), as recorded in a single realisation of the MDM protocol. It reads \cite{Campisi11RMP83}
\begin{align}
       \mathcal{P}(W) = \sum_{m,n}p_n p_{m|n}[U]\delta(W-(E_m-E_n)),
\label{eq:probW}
\end{align}
where $p_n=e^{-\beta E_n}/Z$ is the probability to find the system at energy $E_n$ in the first measurement, and $p_{m|n}[U]$ is the probability of finding the system at energy $E_m$ in the second measurement given that it was found at $E_n$ in the first measurement. Its expression in terms of $U$ and the system Hamitonian eigenprojectors $\Pi_k^s$ (such that $H \Pi_k^s = E_k \Pi_k^s$) reads
\begin{align}
    p_{m|n}[U] = \Tr_S \Pi_m^s U \Pi_n^s U^\dagger \Pi_m^s
    \label{eq:p_m|n[U]}
\end{align}
here $\Tr_S$ denotes the trace operation in the system Hilbert space.

Our approach to experimentally obtain the joint probability $p_{mn}= p_{m|n}[U] p_n $ consists in first imprinting the information about the initial state of our system in an ancillary quantum system (which is a copy of the system itself), let then the system evolve under $U$, and finally measure both ancilla and system energies, so as to obtain, respectively $E_n$ and $E_m$.

The information imprinting stage is achieved by creating the entangled state:
\begin{align}
    \ket{\psi} = \sum_n c_n \ket{n}_a\otimes \ket{n}_s
    \label{eq:psi}
\end{align}
In practice the state $\ket{\psi}$ can be created by preparing the system and ancilla in some factorised state $\ket{\psi_0}_s\otimes \ket{\psi_0}_a$ and then applying an appropriate entangling unitary $V$.
Quite remarkably, by designing the unitary $V$ in such a way that $c_n=\sqrt{p_n}$ one can so simultaneously 
achieve the task of physically preparing the wanted mixed state for the system:
\begin{align}
    \rho_S = \Tr_A \ket{\psi}\bra{\psi} = \sum_n p_n \Pi_n^s
\end{align}
with the populations $p_n =e^{-\beta E_n}/Z$ relative to the system energy eigenstates $\ket{n}_s$ (here $\Tr_A$ denotes trace over the ancilla Hilbert space).

With the preparation in Eq. (\ref{eq:psi}) the joint probability $Q_{mn}$ of measuring $E_n$ in the ancilla and $E_m$ in the system equals the joint probability $p_{mn}=p_n p_{m|n}[U]$, Eq. (\ref{eq:p_m|n[U]}):
\begin{align}
    Q_{mn}  &= \Tr\, (\Pi_m^s U  \otimes \Pi_n^a)
                    \ket{\psi}\bra{\psi}
                    (U^\dagger \Pi_m^s  \otimes \Pi_n^a) \nonumber \\
            &= p_n \Tr\, (\Pi_m^s U \otimes \mathbb{1}_a) 
                    (\Pi_n^s\otimes \Pi_n^a)
                    (U^\dagger \Pi_m^s  \otimes \mathbb{1}_a) \nonumber \\
            &= p_n \Tr_S \Pi_m^s U \Pi_n^s U^\dagger \Pi_m^s = p_{m|n}[U] p_n=p_{mn}\, .
\end{align}
Here $\Pi_n^a$ are the ancilla eigenprojectors, and $\mathbb{1}_a$  denotes the identity operator in the ancilla Hilbert space. In going from the first to the second line we used the salient equation 
\begin{align}
(\mathbb{1}_s \otimes \Pi_n^a) \ket{\psi}= \sqrt{p_n} \ket{n}_s \otimes \ket{n}_a\, ,
\end{align}
 which reflects the essential feature of entanglement. In going from the second to the third line we used the rule $\Tr=\Tr_A\Tr_S$ and assumed the projector operators have unit trace, specifically $\Tr_A \Pi_n^a=1$. 

Note that, as can be seen from the above derivation, since the measurement operators of the ancilla $\mathbb{1}_s\otimes \Pi_n^a$ commute with the operators $\Pi_m^s U_s\otimes \mathbb{1}_a$, whether they are performed before, after or simultaneously with the final system measurement is irrelevant. The information on the initial state of the system is encoded in the ancilla once and for all, and can be retrieved later at any time, without influencing the system.  

We remark that the method does not rely on the fact that the evolution of the main qubit is unitary, hence our derivation continues to hold for a generic evolution described by a generic quantum channel. In the following we shall, in fact demonstrate it both for unitary and non-unitary channels, e.g., when a qubit is subject to projective measurements, or when it interacts with other qubits.

\begin{figure}
    \centering
    \includegraphics[scale=0.5]{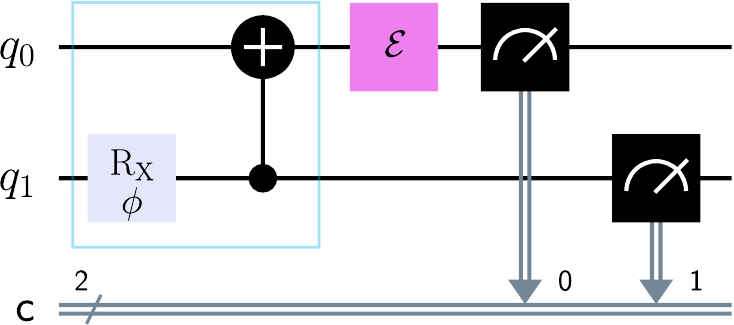}
    \caption{Quantum circuit representation of the AATPM scheme. The two-qubit operation in the blue box creates the purified state, Eq. (\ref{eq:thermalpurification}), that both prepares the main qubit $q_0$ in a thermal mixture, and imprints the information about its state in the ancilla qubit $q_1$. The main qubit evolves according to a generic map $\mathcal{E}$, pink box. Finally, both qubits are read, providing information about the state of the main qubit before and after the evolution.
    }\label{fig:circuit_diagram_2p_Hadamard}
\end{figure}

Figure \ref{fig:circuit_diagram_2p_Hadamard} shows the quantum-circuit representation of the AATPM scheme, for the case of a system being composed by a single qubit with Hamiltonian 
\begin{align}
H_{q_0}=\frac{\omega}{2} \sigma_z^0
\end{align}
with $\omega$ the energy level spacing. 
The qubit of interest is denoted as $q_0$ while $q_1$ is the ancilla. The first block, consisting of a rotation of an angle 
\begin{align}
\phi = 2\arctan(e^{\beta \omega/2}),    
\end{align}
along the X direction of the qubit $q_1$ followed by a CNOT gate prepares the entangled state:
\begin{align}
|\psi\rangle = \frac{1}{\sqrt{2 \cosh(\beta\omega/2)}}(e^{-\beta\omega/4}|0\rangle_0 \otimes |0\rangle_1 + e^{\beta\omega/4}|1\rangle_0 \otimes |1\rangle_1).
\label{eq:thermalpurification}
\end{align}
Here $|k\rangle_{i}$ denotes the energy eigenstates of the system ($i=0$), and ancilla ($i=1$).
Accordingly the qubit of interest is prepared in the Gibbs state 
\begin{align}
\rho_{q_0}= \mathrm{Tr}_{q_0}|\psi\rangle\langle\psi|= \frac{e^{-\beta\omega\sigma^z/2}}{2 \cosh(\beta\omega/2)},
\end{align}
of rescaled inverse temperature 
\begin{align}
\beta\omega =   \ln \tan (\phi/2) .
\label{eq:inv-temp-nominal}
\end{align}

Subsequently it  evolves according to a generic evolution $\mathcal{E}$, and finally both qubits are read in their respective $\sigma_z$ basis. By repeating the circuit $\mathcal{N}$ times one can experimentally determine the probabilities $p_{q_i}^k$ to find qubit $q_i$ in the eigenstate $|k\rangle_i$, and hence measure the joint probability $p_{mn}$ as $p_{mn}=Q_{mn}=p_{q_1}^m p_{q_0}^n$.

\subsection{Experimental validation}

It is well known that quantum systems prepared in a thermal state and then evolving under a unitary driving, satisfy the celebrated Jarzynski identity
\begin{align}
       \langle e^{-\beta W}\rangle = e^{-\beta \Delta F},
       \label{eq:Jarzynski_identity}
\end{align}
where $\beta$ is the initial inverse temperature, $W$ is the work exchanged during the driving protocol, the average $\left\langle...\right\rangle$ is taken with respect to the work statistics defined in Eq. (\ref{eq:probW}), while $-\Delta F$ is the difference between the initial free energy and the free energy the system would have if it was in equilibrium, at inverse temperature $\beta$, at the end of the protocol.

In order to validate our AATMP method we implemented it on IBM quantum processors and used it to study the validity of the Jarzynski equality. In our implementation, after preparation in the thermal state, the qubit of interest $q_0$ was evolved according to a Hadamard gate, and $\Delta F=0$. We collected the qubits statistics over a sample of size $\mathcal{N}=8192$, and repeated the experiments for various vales of the angle $\phi$.

The results were benchmarked against the results obtained by directly implementing the TPM scheme using the newly added IBM capabilities.
\begin{figure}[t]
    \centering
    \includegraphics[width=\linewidth]{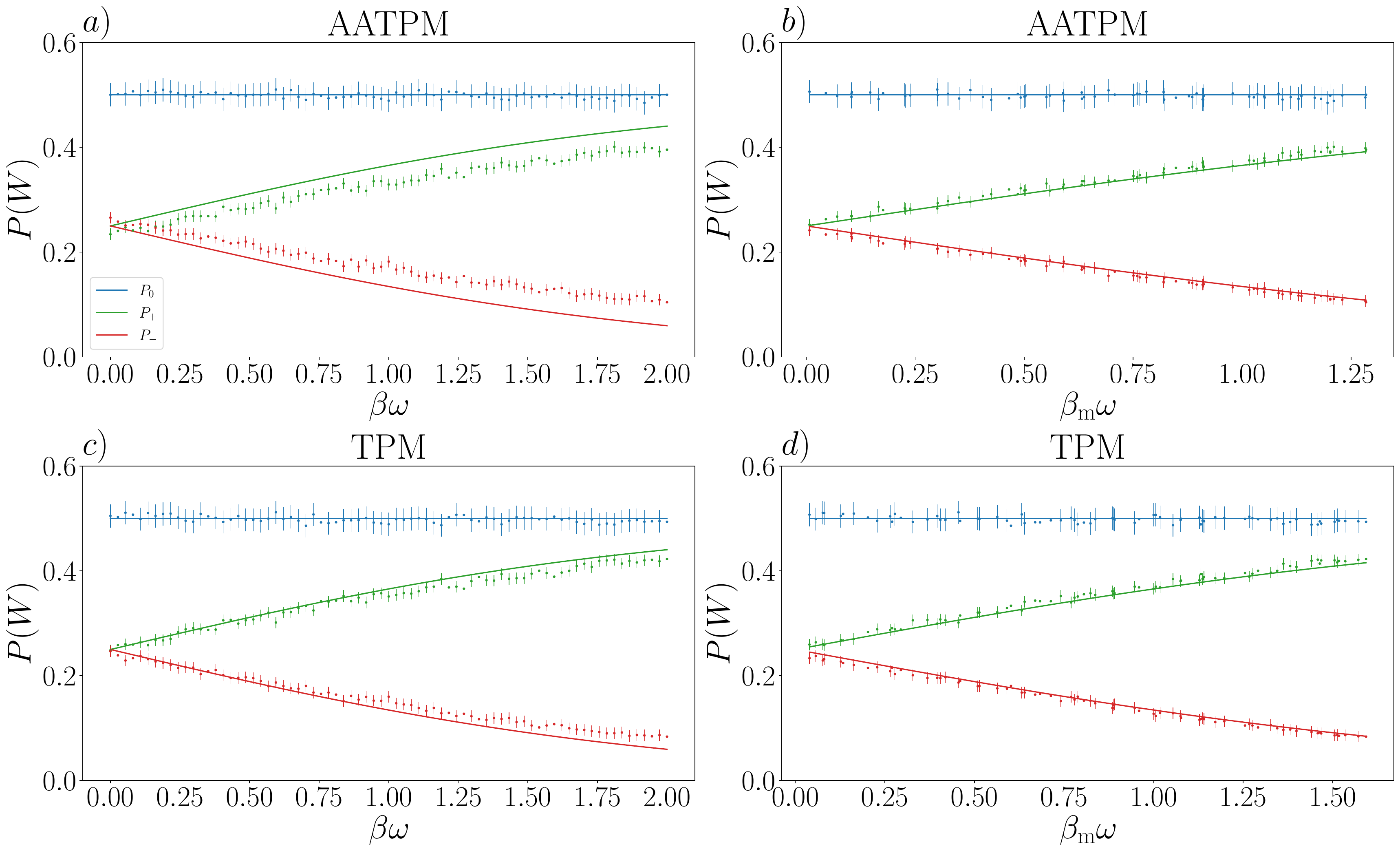}
    \caption{Panels a,c) Work statistics as a function of nominal rescaled inverse temperature for the standard TPM scheme and the AATPM scheme, respectively), for a qubit driven by a Hadamard gate. Panels b,d) same as panels a,c) but as a function of measured rescaled inverse temperature. Dots: experimental data. Solid lines: theory.
    }
    \label{fig:Pw_2p_Hadamard}
\end{figure}

\begin{figure}[t]
    \centering
    \includegraphics[scale=0.285]{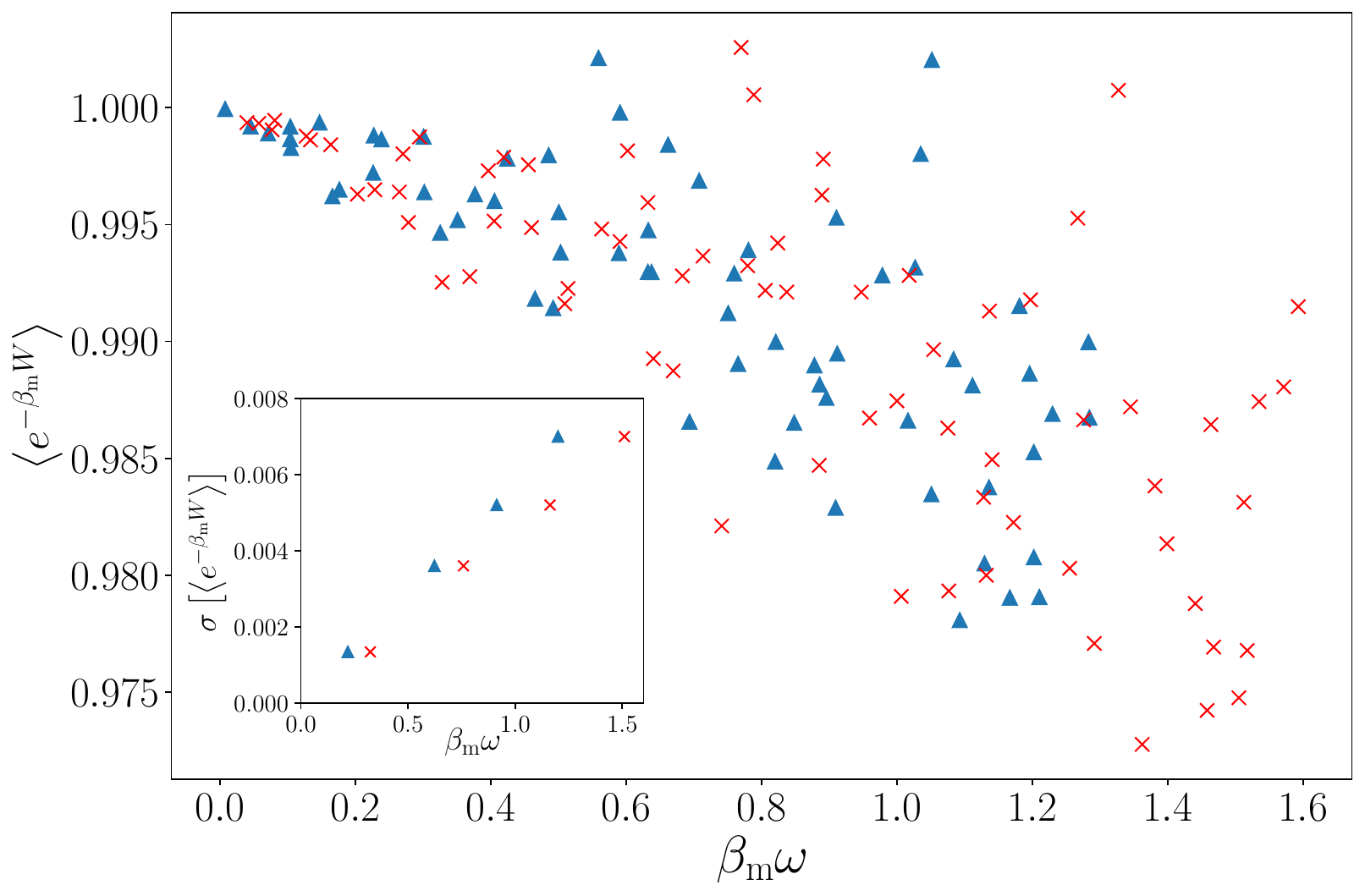}
    \caption{Average exponentiated work $\av{e^{-\beta_\mathrm{m} W}}$ for a qubit driven by a Hadamard gate, as measured using the standard TPM (red crosses) and the AATPM (blue triangles), for various values of measured rescaled inverse temperature $\beta_\mathrm{m}\omega$. Inset, the relative statistical uncertainty for some values of $\beta_\mathrm{m}\omega$ obtained as the standard deviations of averages as detailed in the text.}
    \label{fig:Jarzinsky_2p_Hadamard}
\end{figure}

Figure \ref{fig:Pw_2p_Hadamard}, panel a) shows  the measured values of the probabilities $P_+\doteq Q_{01}$, $P_-\doteq Q_{10}$ and $P_0 \doteq Q_{00}+Q_{11}$, that the qubit undergoes, respectively, the energy changes $ \omega, - \omega, 0$, as function of the rescaled inverse temperature $\beta\omega=\ln \tan (\phi/2)$, Eq. (\ref{eq:inv-temp-nominal}).
The error bars in Fig. \ref{fig:Pw_2p_Hadamard} represent the statistical error due to the finiteness of our samples, and were accordingly calculated as:
$\delta Q_{01}\simeq \delta Q_{10}\simeq 1/\sqrt{\mathcal{N}}$ and $\delta(Q_{00}+Q_{11})\simeq2/\sqrt{\mathcal{N}}$.

Figure \ref{fig:Pw_2p_Hadamard}, panel c), shows the same quantities as in panel a) but for the standard TPM scheme, i.e., obtained by replacing the measurement on the ancilla, with a direct measurement on the qubit of interest, before the application of the driving (Hadamard gate).

Panels a) and c) of Fig. \ref{fig:Pw_2p_Hadamard} evidence a systematic discrepancy between observed and theoretical data,
for both methods.
This is due to a mismatch between the nominal rescaled inverse temperature $\beta\omega$, Eq. (\ref{eq:inv-temp-nominal}) and the actual rescaled inverse temperature acquired by the qubits after the purification step. 
In the AATPM scheme the latter can be read-off the measured populations of the ancilla as
\begin{align}
       \beta_{\mathrm{m}}\omega = \ln [{p^1_{q_1}}/{p^0_{q_1}}]
       \label{eq:measured_temperature}
\end{align}
Similarly, in the standard TPM scheme, they are read off the populations of the first measurement of the qubit of interest.

Panels b) and d) of Fig. \ref{fig:Pw_2p_Hadamard} report the measured probabilities $P_\pm,P_0$ as function of the measured rescaled inverse temperature $\beta_{\mathrm{m}}\omega$. Good agreement with the theoretical expectations is now achieved for both methods. This demonstrates the validity and efficacy of the AATPM scheme.

In Fig. \ref{fig:Jarzinsky_2p_Hadamard} we plot the quantity $\left\langle e^{-\beta_m W} \right\rangle$ for the two methods, blue triangles refer to the AATPM scheme while the red crosses are obtained with the standard TPM scheme. As in Fig. \ref{fig:Jarzinsky_2p_Hadamard}, the statistics is generated for each value of $\beta_\mathrm{m}\omega$ from a sample of size $\mathcal{N}=8192$.
In order to estimate the statistical uncertainty on our estimation of the above quantity we have selected a few values of temperature, have repeated the estimation of $\left\langle e^{-\beta_m W} \right\rangle$ $k=225$ times, and have taken
the standard deviation of the obtained values as measure of the statistical error affecting our data.
Such value of statistical error is reported for 4 values of $\beta_{\mathrm{m}}\omega$ in the inset of Fig. \ref{fig:Jarzinsky_2p_Hadamard}. 
Note how the error tends to increase with increasing $\beta_{\mathrm{m}}\omega$.
Excellent agreement with Eq. (\ref{eq:Jarzynski_identity}) is found for small values of $\beta_\mathrm{m}\omega$. As the temperature decreases ($\beta_{\mathrm{m}}$ increases) the quantity $\left\langle e^{-\beta_m W} \right\rangle$ tends to be underestimated.

This fact may be due to environmental energy dissipation. In such a case, 
the qubit is not only exchanging energy with the work-source during the driving gate in the form of work $W$, but also in the form of heat $Q$ with all other uncontrolled agents. This means that we cannot identify the measured qubit energy change $\Delta E$ as work, $W$, but should identify it as work plus heat $\Delta E=Q+W$.  Then, assuming small heat exchange $\beta Q \ll 1$,
\begin{align}
    \langle e^{-\beta\Delta E}\rangle = \langle e^{-\beta(W+Q)}\rangle =1-\beta\langle Q e^{-\beta W}\rangle +O(\beta Q)^2.
\end{align} 
If this picture is correct, by comparison with the experimental data, it means that $Q$ is positive, namely the qubit gets heated during the experimental run. The above formula, featuring a correction term being (approximately) linear in $\beta$ also provides a possible explanation of the observed increasing error with increasing $\beta_{\mathrm{m}}$.

\begin{figure*}
    \centering
    \includegraphics[scale=0.94]{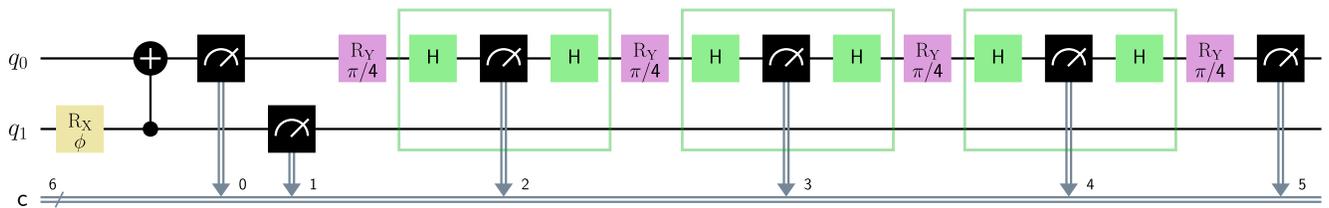}
    \caption{Quantum circuit employed for the experimental validation of the Jarzynski equality Eq. (\ref{eq:Jarzynski_identity}) in presence of $N-1$ intermediate projective measurements (the case $N=4$ is displayed). The green box operations are projective measurements in the $\sigma_x$ basis. The main qubit $q_0$ initial state is both read off directly, according to the standard TPM, and indirectly by reading the state of the ancilla, according to the AATPM scheme. The outcome of the intermediate measurements is ignored.}
    \label{fig:circuit_diagram_projective_measurements}
\end{figure*}

\section{Robustness of fluctuation theorems to intermediate projective measurements}\label{sec:projective_measurements}

With Refs. \cite{Campisi11PRE83, Campisi10PRL105} it has been established that the Jarzynski equality, Eq. (\ref{eq:Jarzynski_identity}), continues to hold true even if, during the driving protocol, an arbitrary number of projective measurements of observables which possibly do not commute with the system Hamiltonian \cite{Campisi11PRE83}, are performed.
Here we experimentally check the validity of that prediction, by taking advantage of the newly added possibility to perform projective measurements at any time during the execution of a quantum circuit, on IBM processors.

The quantum circuit employed to experimentally validate the robustness of the Jarzynski identity to intermediate measurements is depicted in Fig. \ref{fig:circuit_diagram_projective_measurements}. The circuit starts with the creation of the pure state, Eq. (\ref{eq:thermalpurification}), according to the procedure described above.

The qubit of interest $q_0$ is then first measured in the computational basis $\sigma_z$ (that implements the first measurement of the standard TPM scheme), and then undergoes a sequence of identical unitary evolutions $U$, spaced out by  by projcetive measurements of some observable $A$. Specifically, in our experiments, we chose $U$ and $A$ as not commuting with each other:
\begin{align}
    U &=R_Y(\pi/N)=e^{-i \pi \sigma_y/(2N)} \\
    A &= \sigma_x
\end{align}
We remark that the IBM hardware is equipped with the possibility of performing measurements in the $\sigma_z$  basis of each qubit, only. In order to preform a measurement in a different basis one should accordingly first apply the basis-change unitary gate $V$ that maps the $\sigma_z$ eigenbasis onto the wanted measurement basis,  then measure along $\sigma_z$, and finally apply the inverse unitary gate $V^{-1}$ \cite{Buffoni19PRL122}. To see that, note that if $\Pi_k$ are the eigenprojectors associated to the measurement basis which is available in a certain set-up, and $\pi_k$ are the eigenprojectors associated to the wanted measurement basis (i.e., $A = \sum_k a_k \pi_k$, with $a_k$ the eigenvalues of $A$), it is
\begin{align}
    \pi_k = V^{-1} \Pi_k V
    \label{eq:basis-change}
\end{align}
Specifically, the $\sigma_x$ measurement is implemented in the present architecture, by sandwiching the $\sigma_z$ measurement between two Hadamard gates (see green gates in Fig. \ref{fig:circuit_diagram_projective_measurements}).

After the train of intermediate projective measurements both qubit of interest and ancilla are measured in their respective $\sigma_z$ basis to collect the AATPM probabilities $p_{q_i}^k$ as in the previous section.

As discussed in Refs.
\cite{Campisi11PRE83,Campisi10PRL105}, the conditional probability $p_{m|n}$ that qubit $q_0$ is found in state  $|m\rangle$ at the last measurement, provided that it was found in state $|n\rangle$ in the first measurement, reads:
\begin{align}
      p_{m|n} =\sum_{a_1}\dots\sum_{a_{N-1}}p_{m|a_{N-1}}\prod_{k=1}^{N-2}p_{a_{k+1}|a_k}p_{a_1|n},
\end{align} 
where the $p_{a_{k+1}|a_k} = |\langle a_{k+1}|U|a_k\rangle|^2$ are the condition probabilities that the output of the $k+1$-th
measurement is $a_{k+1}$, given that the output of the $k$-th measurement is $a_k$, while $p_{m|a_N} = |\langle m|U|a_N\rangle|^2$ and $p_{a_1|n} = |\langle a_1|U|n\rangle|^2$.

\begin{figure}[t]
    \centering
    \includegraphics[scale=0.35]{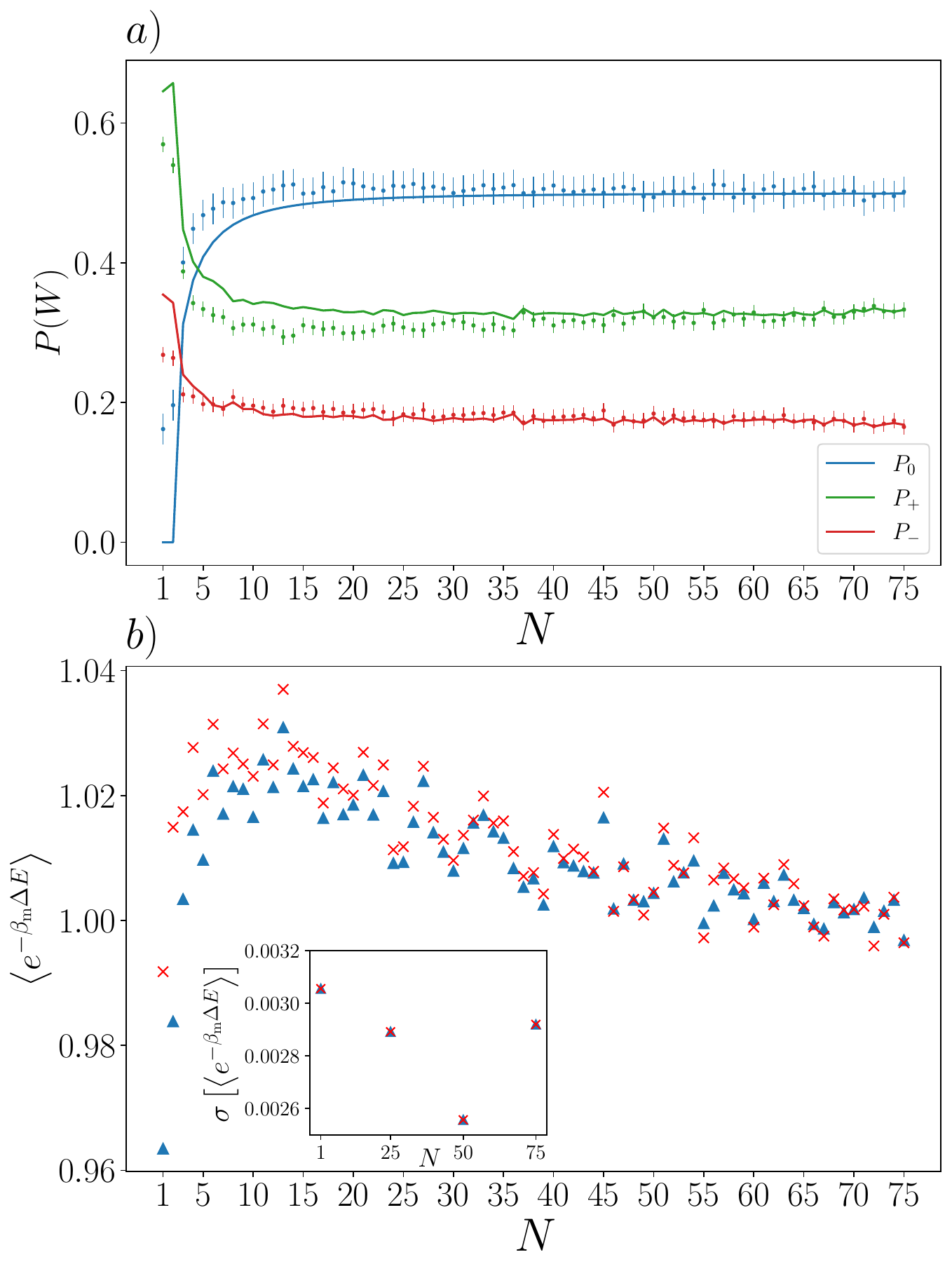}
    \caption{Panel a), work statistics with $N-1$ intermediate projective measurements. Panel b), robustness of the Jarzynski identity in presence of projective measurements, in the inset we show the statistical uncertainty derived as in previous cases for some values of $N=1, 25, 50, 75$. The blue triangles refer to the ancilla assisted method while the red crosses are obtained with the standard two point measurements scheme.}
    \label{fig:Jarzinsky_intermediate_measures}
\end{figure}

With our choice of $U$ and $A$, the transition probabilities $p_{m|n} $, read, for $N>1$,
\begin{subequations}
    \begin{align}
           &p_{0|0} = p_{1|1} = \frac{1}{2}\left(1-\cos^{N-2}\left(\frac{\pi}{N}\right)+\cos^N\left(\frac{\pi}{N}\right)\right)\\
           &p_{0|1} =p_{1|0}=\frac{ \sin^2\left(\frac{\pi}{2N}\right)\left(1+\cos^{N-2}\left(\frac{\pi}{N}\right)\sin^2\left(\frac{\pi}{N}\right)\right)}{\left(1-\cos\left(\frac{\pi}{N}\right)\right)}.
    \end{align}
    \label{eq:pmn_intermediate_measurement}
\end{subequations}
For $N=1$, they read: $p_{0|0} = p_{1|1} = 0$, $p_{0|1} =p_{1|0}= 1$ if $N=1$). 
These analytical expressions are used to compute the values of $P_+,P_-,P_0$ entering the statistics of energy change of the qubit, as in the previous section. Those are plotted for different values of $N$ and fixed nominal inverse temperature $\beta = 1/\omega$
, in Fig. \ref{fig:Jarzinsky_intermediate_measures} panel a), and compared with the values obtained from our experiments. Sample size and errors are evaluated as in the previous section.

Figure \ref{fig:Jarzinsky_intermediate_measures} panel b) shows the quantity  $\langle e^{-\beta_{\mathrm{m}} \Delta E}\rangle $ and its statistical uncertainty (shown in the inset), estimated as in the previous section. Note that we now denote the energy change of the qubit as $\Delta E$, rather than $W$, because now the qubit not only exchanges energy in the form of work (during the unitary gates) but also in the form of so-called ``quantum heat'' (during the measurements) \cite{Elouard17NPJQI3}.
Note that there is a very good agreement between the ancilla assisted results, blue triangles, and those obtained with the standard two point measurement scheme, red crosses. 

We notice that a better agreement is found for larger number of intermediate measurements, for both plots in Fig.  \ref{fig:Jarzinsky_intermediate_measures}. In our understanding this fact has a two-fold origin. On one hand, in the limit of large $N$ it is, in our case, see Eq. (\ref{eq:pmn_intermediate_measurement})
\begin{align}
       \lim_{N\rightarrow\infty}p_{m|n} = \frac{1}{2},\quad\forall m,n.
\end{align}
namely, as a consequence of the protocol applied on the qubit, the latter gets randomised into the completely mixed state $\rho=\mathbb{1}/2$. The phenomenon of randomisation induced by a train of measurements has been observed and discussed earlier, see e.g., \cite{Campisi11PRE83,Yi11PRA84,Giachetti20CM5,Gherardini20arXiv201215216}, but here it occurs as a consequence a different mecahnism. After preparation in a state that is diagonal in the $\sigma_z$ basis, the qubit undergoes a rotation of a vanishingly small angle in the large $N$ limit. That is in the limit of infinite $N$ it remains along $z$, and the subsequent measurement in the $\sigma_x$ basis fully randomise it, leaving in the fully mixed state $\mathbb{1}/2$. The subsequent evolution interrupted by further measurements in the same basis of $\sigma_x$, is frozen due to the quantum Zeno effect \cite{Misra77JMP18,Peres80AJP48}, i.e., the system does not evolve at all, thus remaining at all times in the fully mixed state $\mathbb{1}/2$. As a consequence all correlations between initial  and final energy measurements are lost and the joint probability factorises $p_{mn} = p_n p_m = p_n/2$. Accordingly, the average exponentiated energy change is not subject to the sources of error that typically affect the conditional probability $p_{m|n}$:
\begin{align}
      \langle e^{-\beta\Delta E}\rangle &= \sum_m \frac{e^{-\beta E_m}}{2}  \sum_np_ne^{\beta E_n}
      = \frac{Z}{2}\sum_n\frac{e^{-\beta E_n}}{Z}e^{\beta E_n} = 1.
\end{align}

\begin{figure}
    \centering
    \includegraphics[scale=0.4]{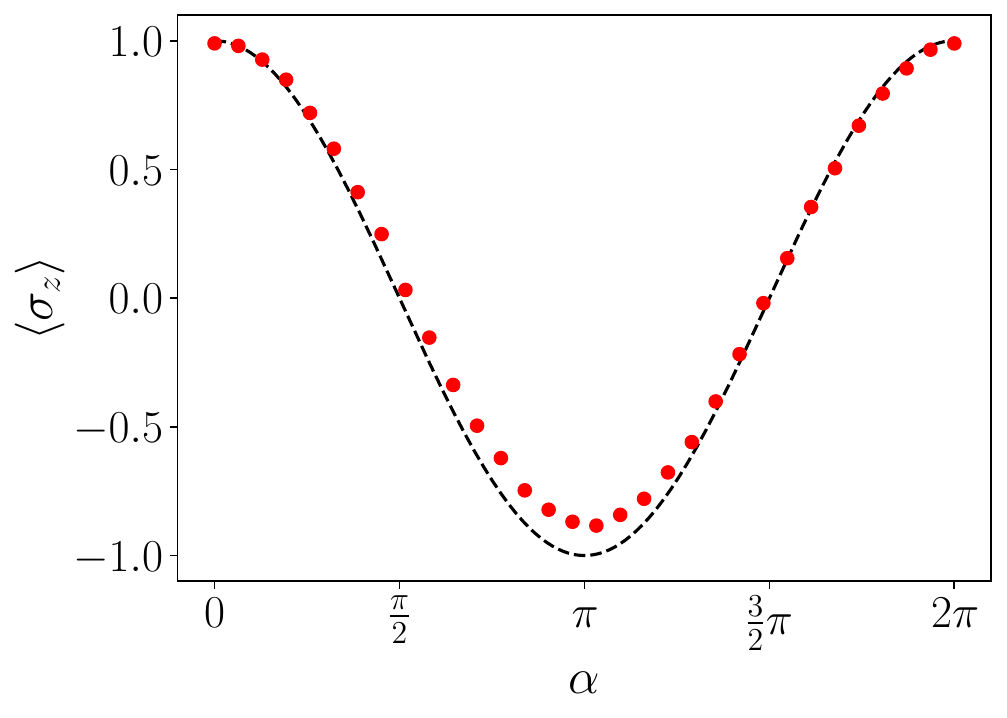}
    \caption{Expectation value of $\sigma_z$ for a single qubit initialised in $|0\rangle$ after the application of a $\mathrm{R_Y}(\alpha)$ gate. Red dots experimental result, dashed line theoretical prediction. }
    \label{fig:qubit_rotation}
\end{figure}

On the other hand, we notice that rotations of angles close to $\pi$ are afftected by larger error, see Fig. \ref{fig:qubit_rotation}. It shows the measured values of $\langle \sigma_z \rangle$ as a function of the rotation angle $\alpha$ against its theoretical expectation, for a single qubit that is initialized in $|0\rangle$ and is rotated according to the gate $\mathrm{R_Y}(\alpha)$, for different values of $\alpha$. The figure evidences that larger deviations are found for angles close $\pi$. In our experiment, the lower is $N$, the closer are rotation angles to $\pi$, hence the larger is the error brought up by each single rotation. This could be another reason why better agreement was observed for large $N$.

\section{Heat engine fluctuation relation}

Any driven bi-partite system subject to a cyclic driving and evolving from tensor product of Gibbs states for each subsystem obeys the following multivariate fluctuation relation \cite{Andrieux09NJP11,Campisi11RMP83,Campisi15NJP17}
\begin{equation}
    \langle e^{-\beta_1\Delta E_1 -\beta_2\Delta E_2} \rangle = 1.
    \label{eq:heat_engine_fluctuation_relation}
\end{equation}
When the work $W$, i.e., the sum of energy changes $\Delta E_1+\Delta E_2$ of the two parts of the system, is not negligible, as in the case of heat engines, the above is often referred to as the heat engine fluctuation relation \cite{Campisi14JPA47,Campisi15NJP17}.
The average $\langle... \rangle$ is now over the joint statistics of the energy exchanges $\Delta E_1$ and $\Delta E_2$, reading
\begin{align}
      &P[\Delta E_1,\Delta E_2] = \sum_{n_1,n_2,m_1,m_2}p_{n_1n_2}p_{m_1m_2|n_1n_2}[U]\nonumber\\
      &\times\delta(\Delta E_1-(E_{m_1}^1-E_{n_1}^1))\delta(\Delta E_2-(E_{m_2}^2-E_{n_2}^2)),
    \label{eq:pDeltaE1DeltaE2}
\end{align}
where $E_{n_i}^i$ are the subsystem $i$ eigenenergies,
\begin{align}
    p_{n_1n_2} = \frac{e^{-(\beta_1 E_{n_1}^1+\beta_2 E_{n_2}^2)}}{Z_1Z_2},
\end{align} 
are the initial populations of the system, and 
\begin{align}
    p_{m_1 m_2|n_1 n_2}[U] = |\langle m_1m_2|U|n_1n_2\rangle|^2,
\end{align}
with $|n_1n_2\rangle$ and the eigenstates of the total Hamiltonian. 

The heat engine fluctuation relation (\ref{eq:heat_engine_fluctuation_relation}) has been experimentally investigated only very recently with an NMR platform (using a quantum Otto heat engine design) by means of interferometric techniques that allow for the measurement of the characteristic function of the statistics $P[\Delta E_1,\Delta E_2]$ \cite{Denzler21arXiv:2104.13427}. Here we further corroborate it by implementing a SWAP quantum heat engine on an IBM quantum processor. We remark that the current AATPM is extremely more effective, in terms of quantum processing time, than the interferometric method, employed in Ref. \cite{Denzler21arXiv:2104.13427}.

We further, for the first time, establish that the heat engine fluctuation relation continues to hold for quantum heat engines that rather than being fuelled by unitary gates, are fuelled by projective measurements, such as the quantum measurement cooling device reported in \cite{Buffoni20QST5}.

\subsection{SWAP quantum heat engine}\label{sec:swap_quantum_heat_engine}
The SWAP quantum heat engine, first introduced in Refs. \cite{Lloyd97PRA56, Quan07PRE76, Allahverdyan08PRE77} is a two-qubit/two-stroke heat engine design that has recently become more and more popular as a prototype of a quantum heat engine, see e.g., \cite{Campisi15NJP17,Timpanaro19PRL123,Uzdin14NJP16}. Despite its broad interest, its experimental implementation has not been reported so far. Using the methods and tools developed above, we have implemented it on the IBM architecture. 
The working substance of a SWAP quantum heat engine is made of two qubits. Let
\begin{align}
H_{q_i} = \frac{\omega_i}{2}\sigma_i^z,\quad i=1,2,\label{eq:qubit_ham}
\end{align}
denote their Hamiltonians, where $\sigma_i^z$ denotes the $z$ Pauli sigma matrix of the $i$-th qubit. At the beginning of the cycle the qubits are decoupled and each at thermal equilibrium at inverse temperatures, $\beta_i, i=1,2$, respectively. Their initial state is then
\begin{align}
      \rho = \frac{e^{-\beta_1H_{q_1}}}{Z_1}\otimes\frac{e^{-\beta_2H_{q_2}}}{Z_2},
\label{eq:SWAPinitialstate}
\end{align}
with $Z_i = \textrm{Tr}_ie^{-\beta_i H_{q_i}}$. In the first step of the cycle, the two
qubits undergo a unitary evolution according to the SWAP gate 
\begin{align}
      U_{\mathrm{SWAP}} = \begin{pmatrix}
           1 &0 &0 &0\\
           0 &0 &1 &0\\
           0 &1 &0 &0\\
           0 &0 &0 &1
      \end{pmatrix}.
\end{align}
In the second step, the qubits interact each with a thermal bath at inverse temperature $\beta_i$, until they thermalize and the initial state $\rho$ is re-established, thus closing the cycle.


\begin{figure}[t]
    \centering
    \includegraphics[scale=0.45]{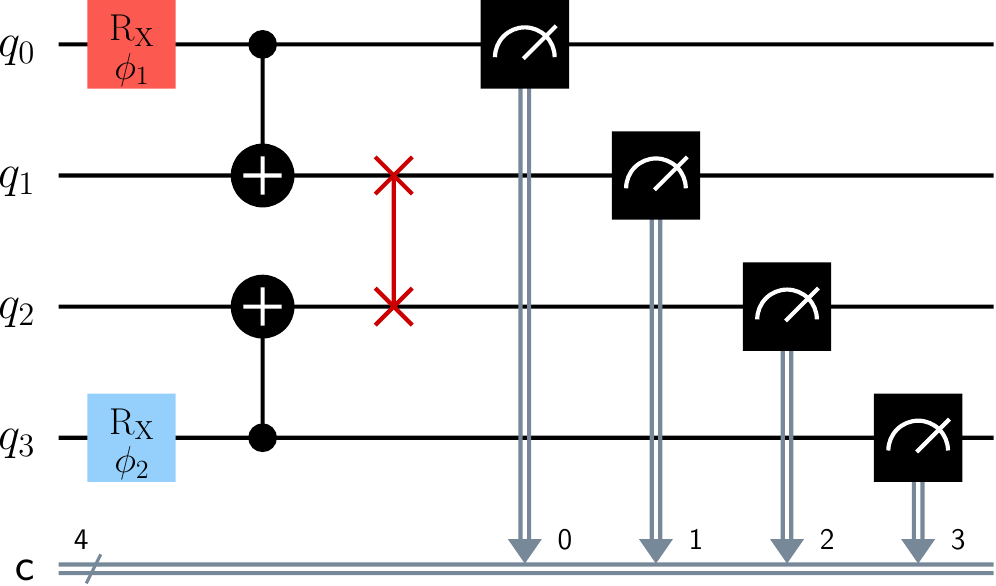}
    \caption{Quantum circuit implementation of the SWAP quantum heat engine. The main qubits $q_1,q_2$ are first prepared in thermal states at different temperatures with the aid of their ancillas $q_0,q_3$ and then undergo a SWAP evolution. Finally their initial states are read-off the ancillas, and their final states are read directly.
    }
    \label{fig:circuit_diagram_swap_engine}
\end{figure}

Fig. \ref{fig:circuit_diagram_swap_engine} shows how we implemented the SWAP engine on the IBM architecture. Qubits $q_1$ and $q_2$ constitute the working substance, while the qubits $q_0$ and $q_3$ are their respective ancillas, which we employ to prepare them in thermal state according to Eq. (\ref{eq:SWAPinitialstate}) and to provide the AATPM capability, as describe above.
We then apply the SWAP gate on the main qubits ($q_1,q_2$), and finally measure all qubits in their respective $\sigma_z$ basis. One could now apply appropriate gates (depending on the outcomes of the measurements) that re-establish the initial state and so close the cycle. For our purposes that is not necessary, since the outcome of the measurements already contains all the information regarding the joint statistics of the energy gained by each qubit, which is sufficient for a full thermodynamic characterisation of the engine 
 \cite{Campisi15NJP17}.


    \begin{figure}[t]
        \centering
        \includegraphics[width=\linewidth]{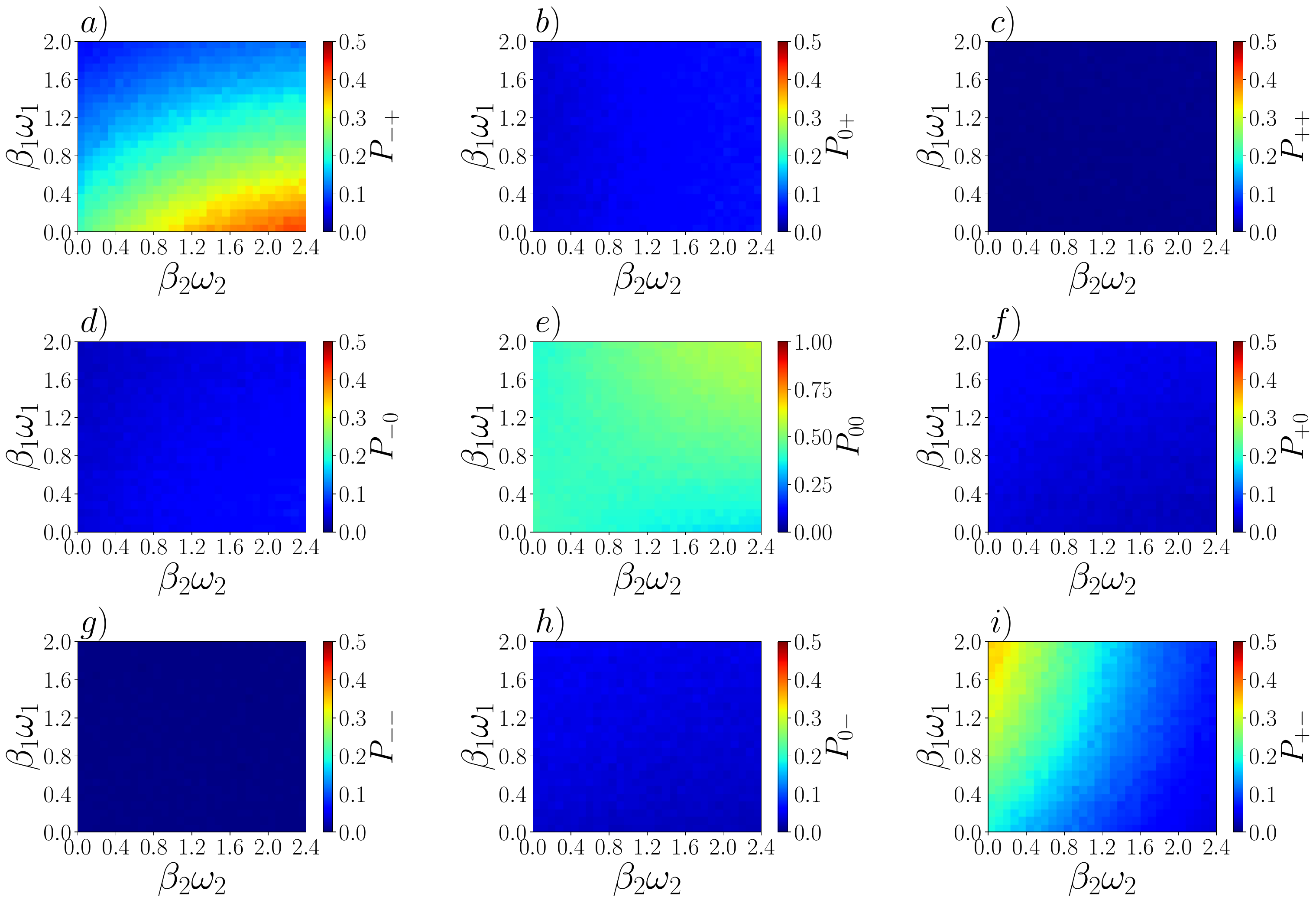}
        \caption{Joint probabilities $P_{ab}$ with $a,b = +,0,-$, Eq. (\ref{eq:P+-0}), for the SWAP engine in Fig. \ref{fig:circuit_diagram_swap_engine}, as a function of the rescaled nominal inverse temperatures  $\beta_1\omega_1$ and $\beta_2\omega_2$.  }
        \label{fig:swap_engine_experimental_joint_probability}
    \end{figure}


For each fixed value of the rotation angles  $\phi_1,\phi_2$, we run the circuit in Fig. \ref{fig:circuit_diagram_swap_engine} $\mathcal{N}=8192$ times, and so collect the probabilities 
$
p_{q_j}^k
$
of finding the qubit $j$ in the state $|k\rangle$ with $j=0,1,2,3$ and $k=0,1$. We recall that $p_{q_0}^k,p_{q_1}^k$ represent the probability that qubit 1 was in state $|k\rangle$ at initial time and final time, respectively. Similarly $p_{q_3}^k,p_{q_2}^k$, represent the probability that qubit 2 was in state $|k\rangle$ at initial time and final time, respectively.
Extending the argument presented in Ref. \ref{sec:AATPM}, we obtain the joint probability   $p_{m_1,m_2,n_1,n_2}=p_{n_1,n_2} p_{m_1 m_2|n_1 n_2}[U]$ as $Q_{m_1,m_2,n_1,n_2}=p_{q_0}^{n_1}p_{q_1}^{m_1}p_{q_2}^{m_2}p_{q_3}^{n_2}$.

Fig. \ref{fig:swap_engine_experimental_joint_probability} shows the probabilities $P_{ab}$, $a,b=-,0,+$, that the energy change of qubit $q_1$ is $\Delta E_1 = a \omega_1$, and the energy change of qubit $q_2$ is $\Delta E_2 = b \omega_2$. 
Those are obtained from the joint probabilities $p_{m_1,m_2,n_1,n_2}$ via the expression
\begin{align}
P_{ab} = \sum_{m_1,m_2,n_1,n_2} p_{m_1,m_2,n_1,n_2} \delta_{m_1-n_1,a} \delta_{m_2-n_2,b}
\end{align}
with $\delta_{i,j}$ the Kronecker symbol.
For all values of nominal rescaled inverse temperatures $\omega_i\beta_i$, we found good agreement between the experimentally measured values and the theoretical ones
\begin{subequations}
       \label{eq:P+-0}
\begin{align}
       &P_{00}=\frac{2\cosh(\beta_1\omega_1/2+\beta_2\omega_2/2)}{Z_1Z_2},\\
       &P_{-+}=\frac{e^{-(\beta_1\omega_1-\beta_2\omega_2)/2}}{Z_1Z_2},\\
       &P_{+-}=\frac{e^{(\beta_1\omega_1/2-\beta_2\omega_2)/2}}{Z_1Z_2},\\
       &P_{\pm 0} = P_{0\pm} = P_ {\pm\pm} = 0,
\end{align}
\end{subequations}
where, for $\beta_i\omega_i$, we used the measured values 
\begin{align}
\beta_{1,m}\omega_1&=\ln[p_1^{q_0}/p_0^{q_0}]\\
\beta_{2,m}\omega_2&=\ln[p_1^{q_3}/p_0^{q_3}].
\end{align}
which, we recall, do not perfectly coincide with the nominal rescaled inverse temperatures 
$
\beta_i\omega_i= \ln \tan (\phi_i/2)
$.
The difference between theoretical values and experimental values of the $P_{ab}$'s was always below the value $0.1$

\begin{figure}[t]
    \centering
    \includegraphics[scale=0.375]{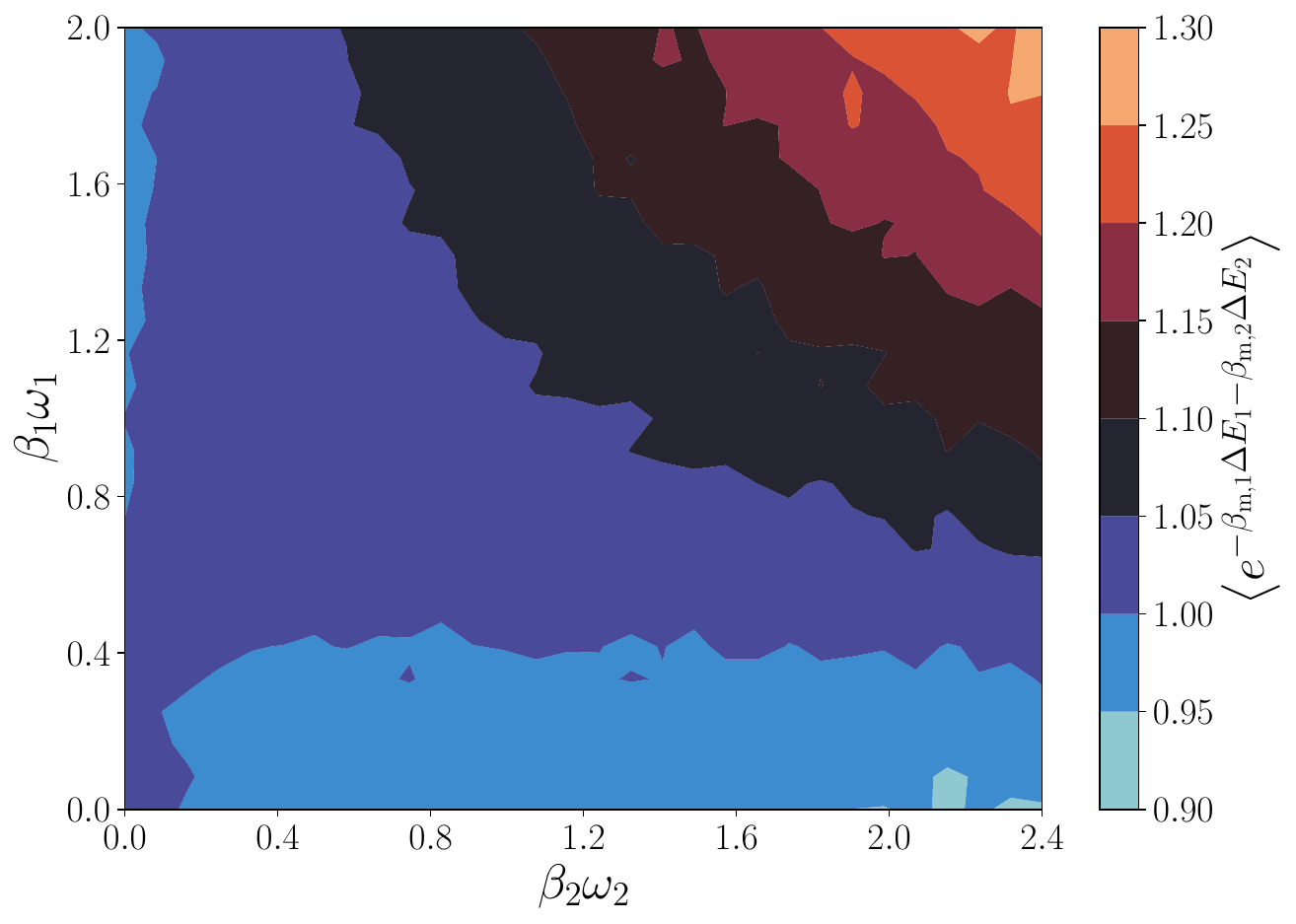}
    \caption{Measured values of $\av{e^{-\beta_\mathrm{m,1}\Delta E_1-\beta_\mathrm{m,2}\Delta E_2}}$ for the SWAP engine in Fig. \ref{fig:circuit_diagram_swap_engine} as a function of nominal inverse rescaled temperatures $\beta_1\omega_1$, $\beta_2\omega_2$.}
    \label{fig:Swap_engine_Jarzinsky}
\end{figure}

Fig. \ref{fig:Swap_engine_Jarzinsky} shows the quantity $\displaystyle\langle e^{-\beta_{\mathrm{m},1} \Delta E_1-\beta_{\mathrm{m},2} \Delta E_2 }\rangle$ as a function of $\beta_1\omega_1$, $\beta_2\omega_2$.
In our implementations the nominal values of the qubits frequencies were $\omega_1 \approx 5.25\,\hbar \mathrm{GHz}$ and $\omega_2 \approx 5.17\,\hbar \mathrm{GHz}$. We notice that the agreement with Eq. (\ref{eq:heat_engine_fluctuation_relation}) is very good for sufficiently small values of $\beta_i\omega_i$ while the deviation from the theoretical value increases as  $\beta_i\omega_i$ increase.
As discussed above, this could be a signature of possible leaks of energy from the environment that heat up the qubits.

\begin{figure}[t]
    \centering
    \includegraphics[width=\linewidth]{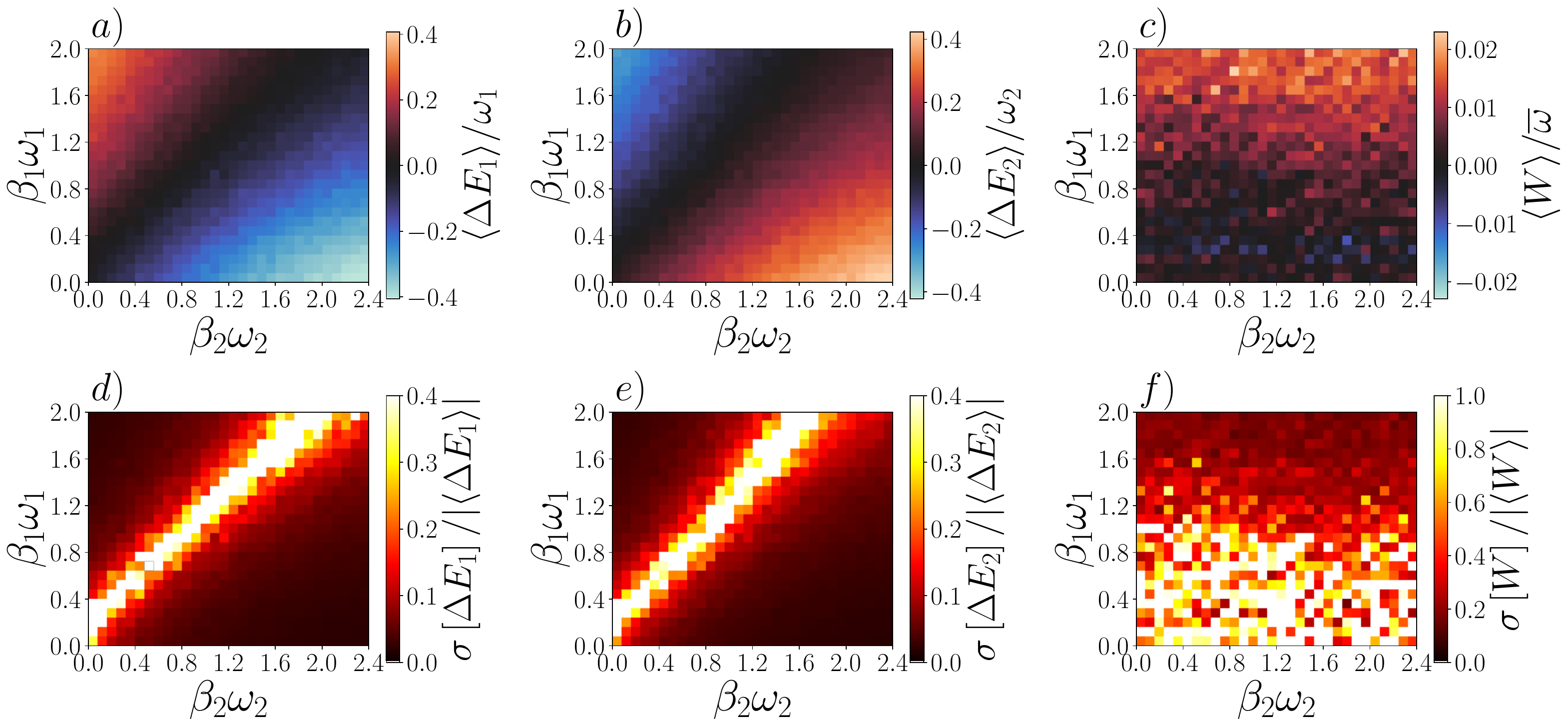}
    \caption{Panels $a)$, $b)$, $c)$: Measured rescaled energy exchanges $\av{\Delta E_1}/\omega_1$, $\av{\Delta E_2}/\omega_2$, $\av{W}/\overline{\omega}$,  for the SWAP engine in Fig. \ref{fig:circuit_diagram_swap_engine} as a function of nominal inverse rescaled temperatures $\beta_1\omega_1$, $\beta_2\omega_2$. Panels $d)$, $e)$, $f)$: their relative experimental error.}
    \label{fig:Swap_engine_E1_E2_W}
\end{figure}

For an ideal SWAP engine, the average energy exchanges and the average work read  \cite{Campisi15NJP17}:
\begin{subequations}
\begin{align}
&\langle\Delta E_1\rangle = \left(\frac{1}{1+e^{\beta_2\omega_2}}-\frac{1}{1+e^{\beta_1\omega_1}}\right)\omega_1,\\
&\langle\Delta E_2\rangle = -\left(\frac{1}{1+e^{\beta_2\omega_2}}-\frac{1}{1+e^{\beta_1\omega_1}}\right)\omega_2,\\
&W = \left(\frac{1}{1+e^{\beta_2\omega_2}}-\frac{1}{1+e^{\beta_1\omega_1}}\right)(\omega_1-\omega_2).
\end{align}\label{eq:swap_energy_exchanges}
\end{subequations}

Fig. \ref{fig:Swap_engine_E1_E2_W}, panels $a)$, $b)$ and $c)$ shows the values of $\langle\Delta E_1\rangle/\omega_1$, $\langle\Delta E_2\rangle/\omega_2$, and $\av{W}/\overline{\omega}$ (where $\overline{\omega}=(\omega_1+\omega_2)/2$) as obtained from the measured populations $p_{q_j}^k$:
\begin{subequations}
    \begin{align}
        &\langle \Delta E_1 \rangle/\omega_1 =  \left[\left( p_{q_1}^1-p_{q_1}^0\right) - \left( p_{q_0}^1-p_{q_0}^0\right) \right]/2,\\
        &\langle \Delta E_2 \rangle/\omega_2 = \left[\left( p_{q_2}^1-p_{q_2}^0\right) - \left( p_{q_3}^1-p_{q_3}^0\right) \right]/2,\\
        &\langle W \rangle/\overline{\omega} = (\langle \Delta E_1 \rangle +\langle \Delta E_2 \rangle)/\overline{\omega},
    \end{align}\label{eqs:swap_exp_energy_exchanges}
\end{subequations}

Fig. \ref{fig:Swap_engine_E1_E2_W}, panels $d)$, $e)$, and $f)$ show their relative fluctuations estimated as
\begin{align}
        &\sigma\left[\langle \Delta E_j \rangle\right] = \frac{2\omega_j}{\sqrt{\mathcal{N}}} ,\quad \sigma\left[\langle W\rangle \right] =2\frac{\omega_1+\omega_2}{\sqrt{\mathcal{N}}},
\end{align}
with $\mathcal{N}=8192$ the size of our statistical sample.

Note that for qubits with same resonant frequency, the expected average work is null, see Eq. (\ref{eq:swap_energy_exchanges}c). In our case the two qubits resonant frequencies were very close, and in fact the according measured work was always very small (and almost always positive), compared to the according resonant frequencies energies, see Fig. \ref{fig:Swap_engine_E1_E2_W}c).

We recall that only four modes of operations are allowed for a generic bi-partite heat engine, depending on the sign of the energy exchanges $\av{\Delta E_i}$ and their sum, i.e., the work $\av{W}$ \cite{Solfanelli20PRB101}. Those are: Refrigerator $[R]$, i.e., the cold subsystem gets colder, the hot subsystem gets hotter while work is injected in the system; Heat Engine $[E]$, i.e., the cold subsystem heats-up, the hot subsystem cools down while work is extracted; Thermal Accelerator $[A]$, i.e., work is spent to heat up the cold subsystem and cool down the hot subsystem; Heater $[H]$, i.e., both subsystems receive energy from the work source. 

\begin{figure}[t]
	\centering
	\includegraphics[width=0.49\linewidth]{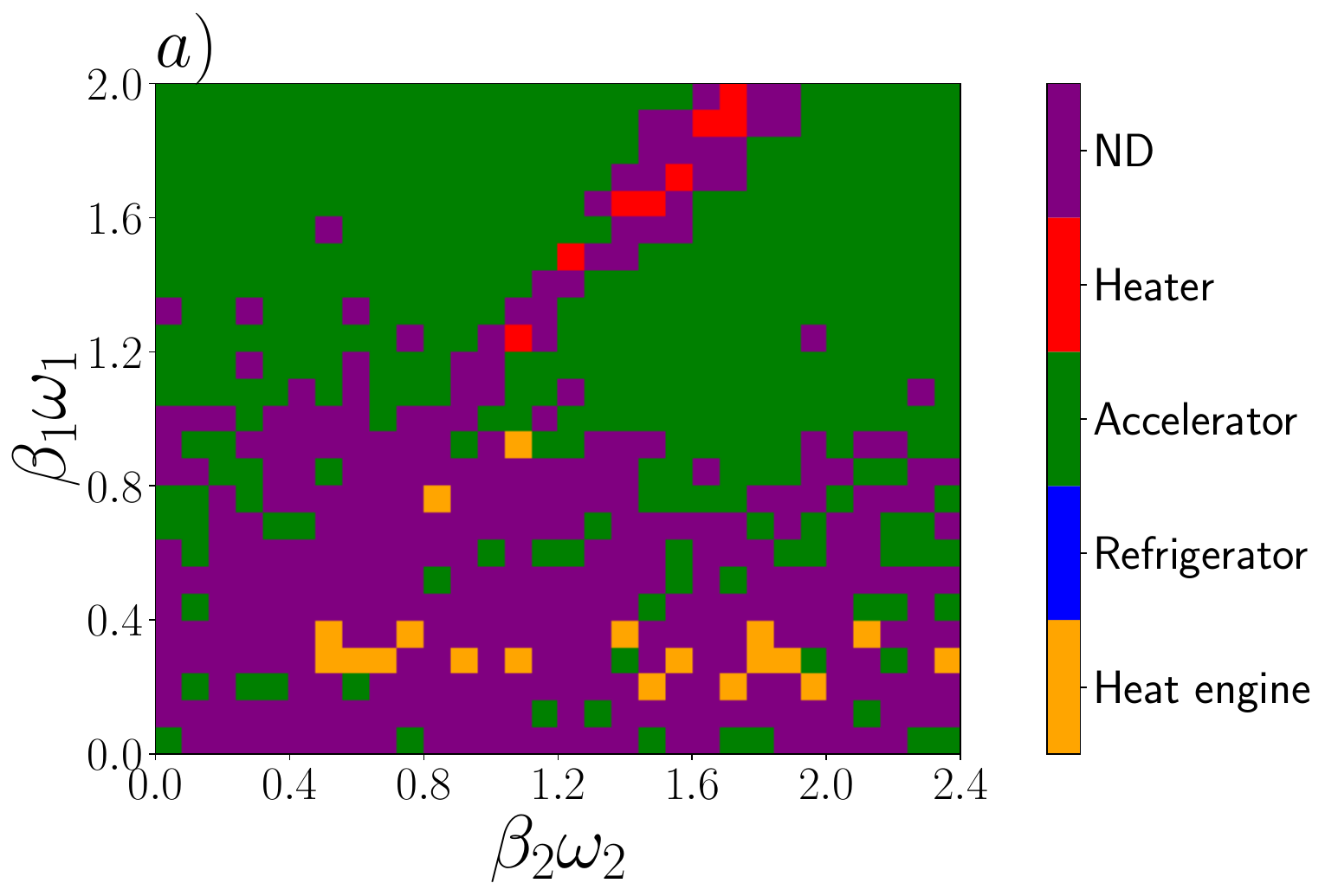}
	\includegraphics[width=0.49\linewidth]{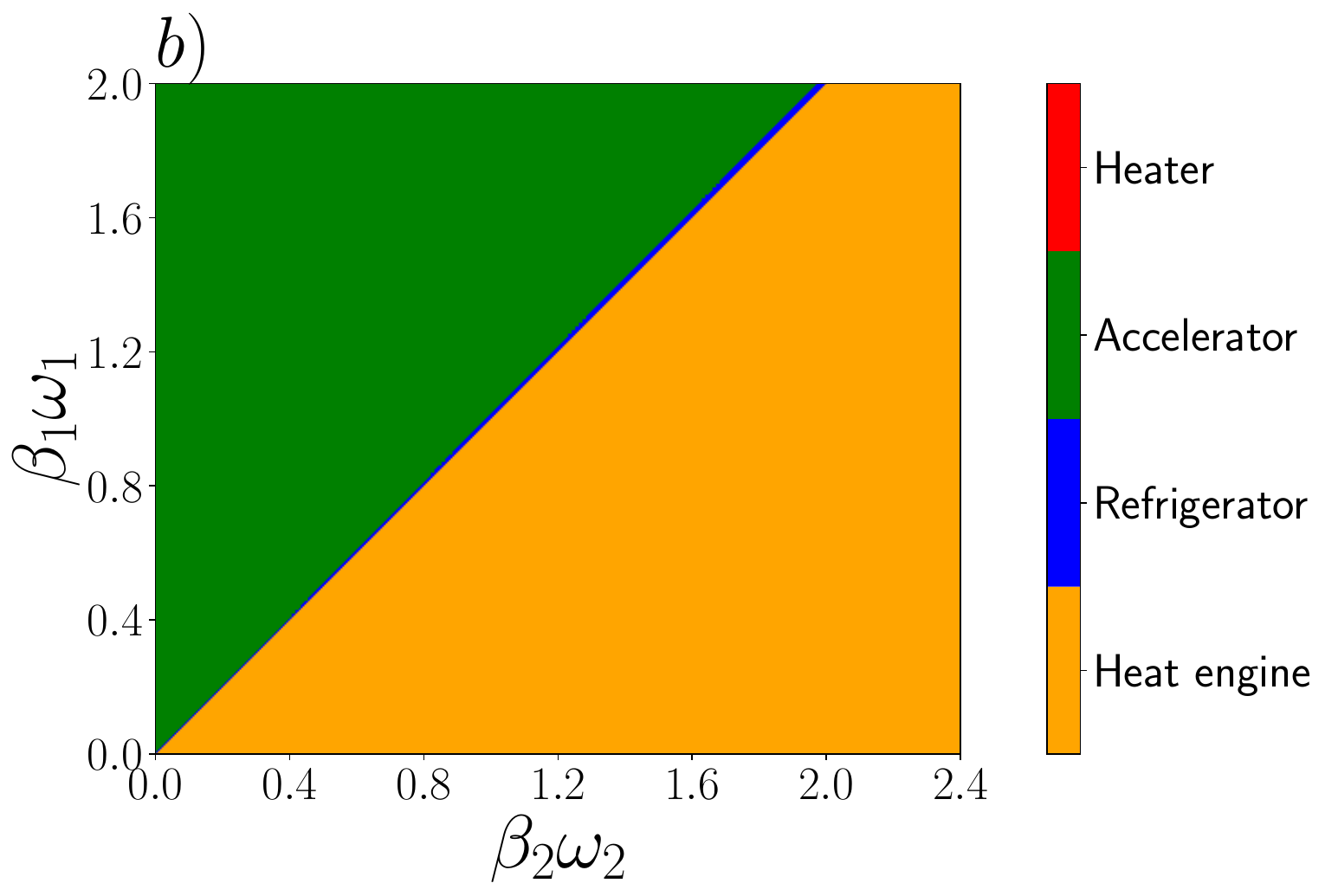}
	\caption{Panel $a)$: experimentally observed modes of operation of the SWAP engine. Panel $b)$: theoretical prediction.}
	\label{fig:Swap_engine_regimes}
\end{figure}

Fig. \ref{fig:Swap_engine_regimes}, panel $a)$ shows the ``phase diagram'' of the experimentally observed modes of operation of the engine as a function of the rescaled inverse nominal temperatures of the two qubits. Panel $b)$ shows the theoretical diagram of the SWAP engine, according to the rule \cite{Campisi15NJP17}:
\begin{subequations}
    \begin{align}
    &[E]:\frac{\beta_1}{\beta_2}\leq\frac{\omega_2}{\omega_1}\\
    &[R]: 0\leq\frac{\omega_2}{\omega_1}\leq\frac{\beta_1}{\beta_2} \label{eq:ideal-regime-R}\\ 
    &[A]: 1\leq\frac{\omega_2}{\omega_1},
    \end{align}
\end{subequations}
We mostly observed the thermal accelerator [A] mode, and sporadically the heat engine [E] and heater mode [H], which accordingly cannot be considered as a robust operations (purple pixel denote cases where the error is large enough that the sign of the interested quantities is not well defined). The refrigerator mode [R] was not observed. This is because the region where we should be able to see it is too thin, see blue region in Fig. \ref{fig:Swap_engine_regimes}b). In order to robustly implement [R], it is crucial to have qubits with appreciably different level spacings, so as to widen the extension of the [R] region. For example, in order to cool down a qubit, say, qubit $q_2$, being at temperature $1/\beta_2$ it is necessary that the other qubit (that must be prepared at higher temperature $1/\beta_1>1/\beta_2$) has a larger level spacing $\omega_1$, see eq. (\ref{eq:ideal-regime-R}).

\subsection{Quantum Measurement Cooling}\label{sec:QMC}

Quantum measurement cooling (QMC) is a cooling paradigm, recently put forward in Ref. \cite{Buffoni19PRL122}, whereby a quantum refrigerator is powered by the very act of quantum measurement, by exploiting its invasiveness.
As discussed in Ref. \cite{Buffoni19PRL122}, such a measurement-powered cooler can be implemented by substituting, in the SWAP engine design described above, the SWAP gate with a projective measurement on a properly chosen basis. As shown in Ref. \cite{Buffoni19PRL122,Solfanelli19JS9} maximal cooling power (heat extracted from the cold qubit per cycle) and maximal thermodynamic cooling efficiency are achieved when the measurement basis is the singlet-triplet basis
    \begin{align}
     \begin{cases}
        |\psi_1\rangle = |00\rangle\\
        |\psi_2\rangle = \displaystyle\frac{|01\rangle+|10\rangle}{\sqrt{2}}\\
        |\psi_3\rangle = \displaystyle\frac{|01\rangle-|10\rangle}{\sqrt{2}}\\
        |\psi_4\rangle = |11\rangle
     \end{cases}
      \label{eq:optimal_basis_qmc}
    \end{align}

\begin{figure}[t]
    \centering
    \includegraphics[width=\linewidth]{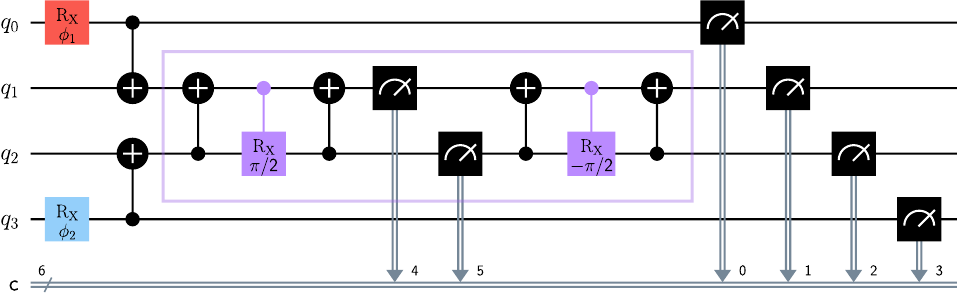}
    \caption{ Quantum circuit implementation of the quantum measurement cooling heat engine design. The main qubits $q_1,q_2$ are first prepared in thermal states at different temperatures with the aid of their ancillas $q_0,q_3$ and then undergo a measurement in the singlet-triplet basis, Eq. (\ref{eq:optimal_basis_qmc}), purple box. Finally their initial states are read-off the ancillas, and their final states are read directly. The outcome of the singlet-triplet basis measurement is discarded.}
    \label{fig:circuit_diagram_quantum_measurment_cooling}
\end{figure}

Figure \ref{fig:circuit_diagram_quantum_measurment_cooling} shows our implementation of QMC on the IBM architecture. The quantum circuit begins with the ancilla assisted preparation of the state (\ref{eq:SWAPinitialstate}). Then a measurement on the singlet-triplet basis is performed according to the basis-change method describe above, Eq. (\ref{eq:basis-change}). In this case the basis change unitary reads
\begin{align}
    V = \begin{pmatrix}
             1 &0 &0 &0\\
             0 &1/2 &1/2 &0\\
             0 &1/2 &1/2 &0\\
             0 &0 &0 &1
        \end{pmatrix}
      =\mathrm{CNOT}\cdot \mathrm{CR}_\mathrm{Y}\left(-\frac{\pi}{2}\right)\cdot \mathrm{CNOT} 
\end{align}
where $\mathrm{CR}_\mathrm{Y}$ is a controlled rotation along the $\mathrm{Y}$-axis.
Finally all qubits are measured in their relative $\sigma_z$ basis, to give the probabilities $p_{q_i}^k$, and hence the full joint statistics of energy exchanges, as in the SWAP case.

\begin{figure}[t]
    \centering
    \includegraphics[scale=0.35]{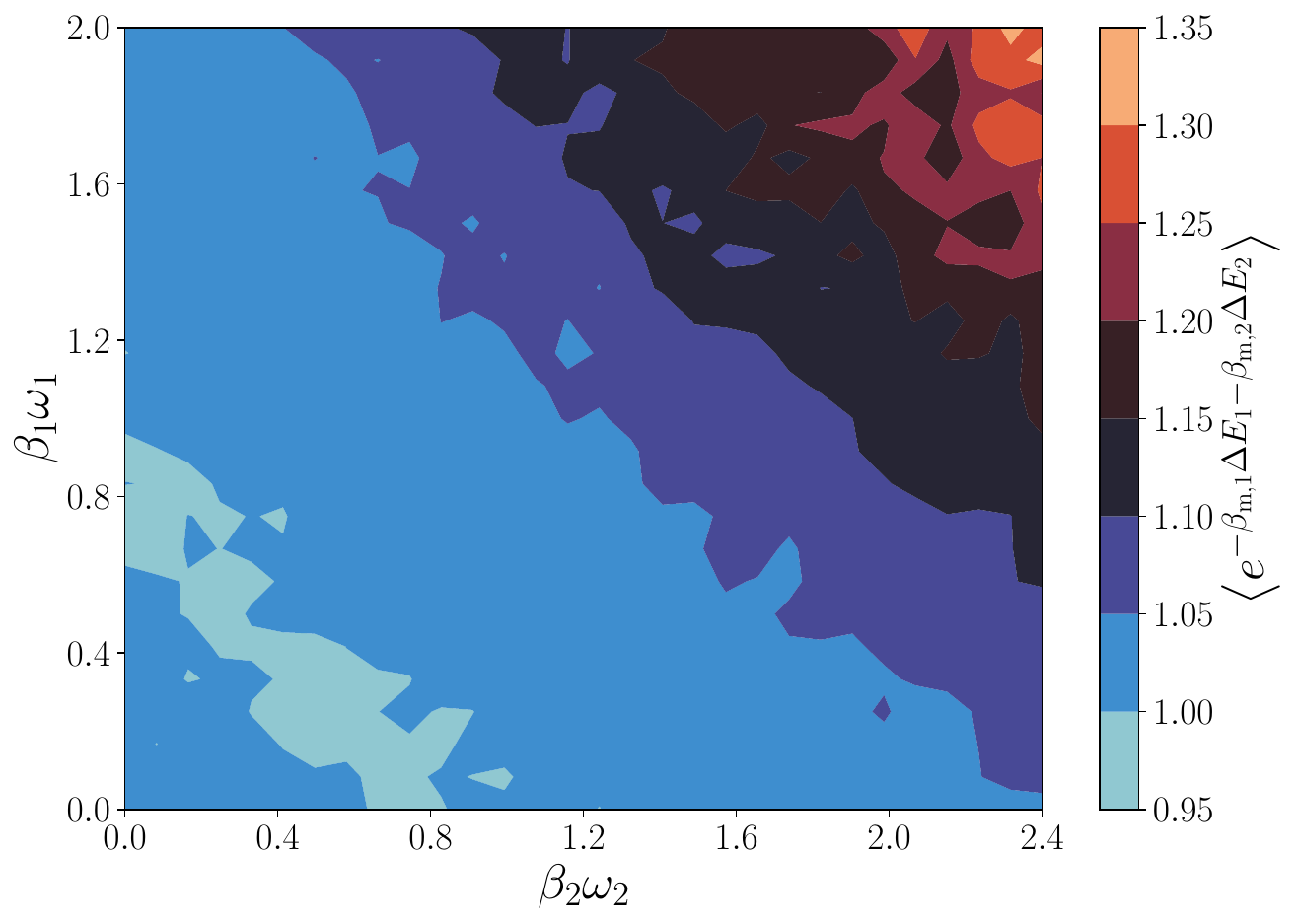}
    \caption{Measured values of $\av{e^{-\beta_\mathrm{m,1}\Delta E_1-\beta_\mathrm{m,2}\Delta E_2}}$  for the QMC engine design in Fig. \ref{fig:circuit_diagram_quantum_measurment_cooling} as a function of nominal inverse rescaled temperatures $\beta_1\omega_1$, $\beta_2\omega_2$.}
    \label{fig:QMC_engine_Jarzinsky}
\end{figure}

We recall that theory predicts that the heat engine fluctuation relation, Eq. (\ref{eq:heat_engine_fluctuation_relation}), is robust to projective measurements \cite{Campisi10PRL105}, thus should continue to hold in the present case. Figure (\ref{fig:QMC_engine_Jarzinsky}) shows the quantity $\langle e^{-\beta_{m,1}\Delta E_1 -\beta_{m,2}\Delta E_2} \rangle$ as a function of $\beta_{i}\omega_i$. We recall that the two qubits resonant frequencies were $\omega_1 \approx 5.25 \mathrm{GHz}\,\hbar$ and $\omega_2 \approx 5.17 \mathrm{GHz}\,\hbar$. As in the SWAP engine setting, we observe very good agreement for small values of $\beta_i\omega_i$, and growing, but still moderate, deviations when those quantities grow. 

\underline{}

Theory predicts that, as compared to the SWAP engine, in this case all energetic exchanges are halved \cite{Buffoni19PRL122},
\begin{subequations}
\begin{align}
&\langle\Delta E_1\rangle = \left(\frac{1}{1+e^{\beta_2\omega_2}}-\frac{1}{1+e^{\beta_1\omega_1}}\right)\frac{\omega_1}{2,}\\
&\langle\Delta E_2\rangle = -\left(\frac{1}{1+e^{\beta_2\omega_2}}-\frac{1}{1+e^{\beta_1\omega_1}}\right)\frac{\omega_2}{2},\\
&W = \left(\frac{1}{1+e^{\beta_2\omega_2}}-\frac{1}{1+e^{\beta_1\omega_1}}\right)\frac{\omega_1-\omega_2}{2}.
\end{align}\label{eq:QMC_energy_exchanges}
\end{subequations}
Accordingly the ``phase diagram'' is identical to that of a SWAP engine.

\begin{figure}[t]
    \centering
    \includegraphics[width=\linewidth]{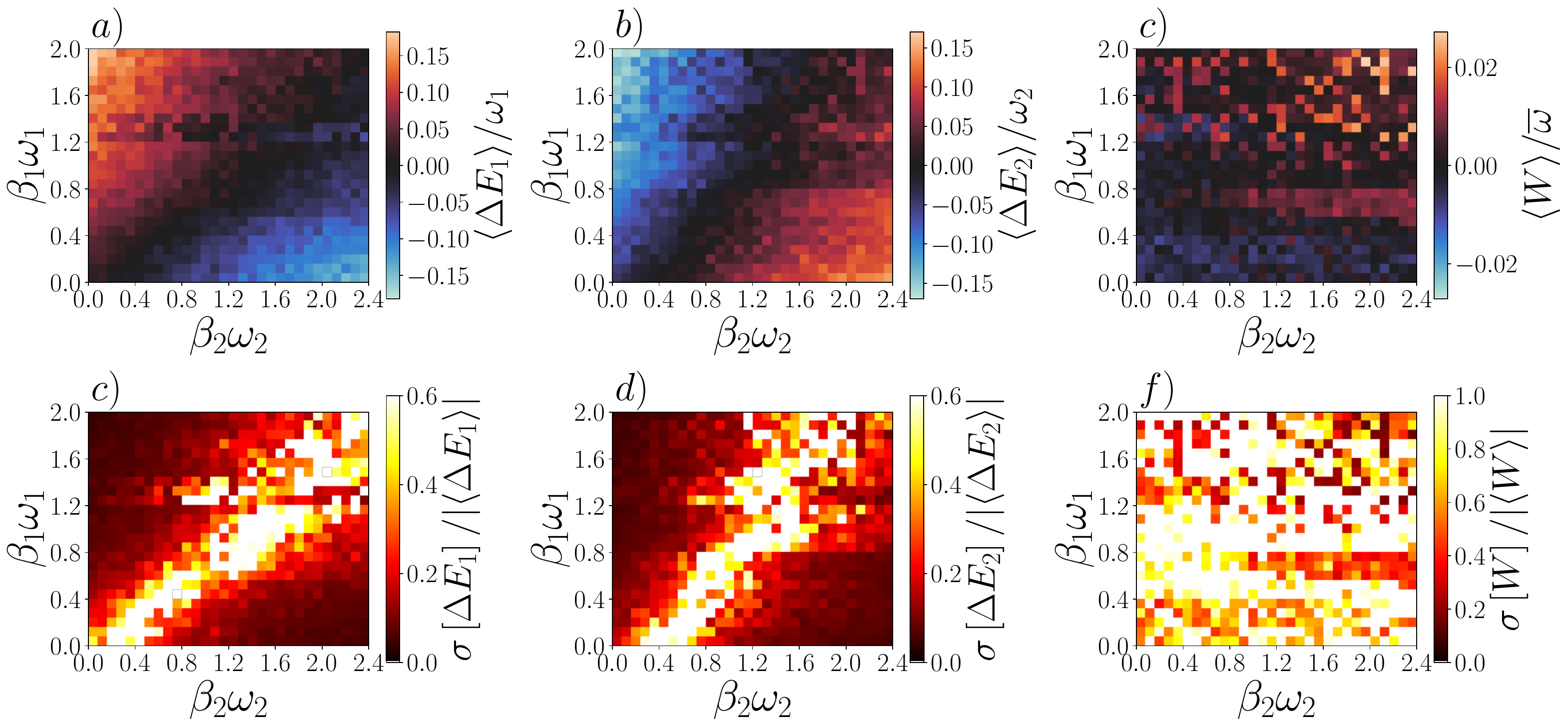}
    \caption{Panels $a)$, $b)$, $c)$: Measured rescaled energy exchanges $\av{\Delta E_1}/\omega_1$, $\av{\Delta E_2}/\omega_2$, $\av{W}/\overline{\omega}$, for the QMC engine design in Fig. \ref{fig:circuit_diagram_quantum_measurment_cooling} as a function of nominal inverse rescaled temperatures $\beta_1\omega_1$, $\beta_2\omega_2$. Panels $d)$, $e)$, $f)$: their relative experimental error.}
    \label{fig:QMC_E1_E2_W}
\end{figure}

Figure \ref{fig:QMC_E1_E2_W}, panels $a)$, $b)$, $c)$, shows the experimental values of $\langle \Delta E_1\rangle/\omega_1$, $\langle \Delta E_2\rangle/\omega_2$ and $\av{W}/\overline{\omega}$. Figure \ref{fig:QMC_E1_E2_W}, panels $d)$, $e)$, $f)$ shows, their respective relative uncertainty, computed as in Eq. (\ref{eqs:swap_exp_energy_exchanges}). Note that the measured values are smaller than half of the according values measured for the SWAP engine. In our understanding that is due to the considerably more complex circuit used to implement this design, which is expected to be affected by larger noise and energy leaks. That this is in fact the case, can be seen by noticing that relative errors are larger in this case compared to the SWAP engine case.

As in the SWAP case, due to the closeness of the qubits level spacings, cooling was not achieved, and the operation observed, for all values of the nominal inverse rescaled temperatures  $\beta_i\omega_i$ was mostly that of a thermal accelerator. The experimentally observed ``phase diagram'' is qualitatively similar to that observed for the SWAP engine, only more noisy. Exactly as in the SWAP engine case, for the implementation of cooling, qubits with appreciably different level spacing is necessary. That however is not a knob that a remote user of the quantum processor can access.

\section{Conclusions}
With this work we have laid the experimental foundations for the study of non-equilibrium thermodynamic phenomena in quantum processors. We have put forward a novel ancilla assisted scheme for the practical implementation of the two point measurement scheme and have experimentally validated its efficacy on an IBM quantum processor. The method can be straightforwardly ported on other quantum computing platforms. Compared to previous methods (specifically the interferometric method \cite{Mazzola13PRL110,Dorner13PRL110}) it is extremely more effective in terms of QPU time necessary for its implementation, and also allows for the preparation of physical thermal states, a problem which is often avoided by mimicking such states.

We implemented the method on IBM quantum processors, and employed it to experimentally corroborate the prediction that the fluctuation theorem is robust to intermediate projective measurements of a generic observable \cite{Campisi10PRL105,Campisi11PRE83}, which was not experimentally observed so far.

We also demonstrated, for the first time, the implementation of a quantum heat engine design  that is in the limelight of current research, namely the two-qubit/two stroke SWAP quantum engine. We employed our ancilla assisted two point measurement scheme to corroborate the validity of the heat engine fluctuation theorem, a theoretical prediction that was so far only addressed with the interferometric method \cite{Denzler21arXiv:2104.13427}. 
We further verified, for the first time, the validity of the  heat engine fluctuation theorem for the case of a measurement fuelled quantum heat engine. That result is at the basis of the mechanism of quantum measurement cooling \cite{Buffoni19PRL122}, whereby a qubit is cooled down by the very act of being measured (along with a second hotter qubit) in an appropriate measurement basis. In practice that design is obtained by replacing the SWAP operation of the SWAP engine, with a measurement on the two-qubit singlet/triplet basis.

For both heat engines implemented, the observed mode of operation was mostly that of a thermal accelerator, namely, physically what happened is that the cold qubit was heated, the hot qubit was cooled down, while energy was spent to make this happen. Interestingly, the same was observed in an experimental study of the thermodynamics of a quantum annealer \cite{Buffoni20QST5}. The fact that other modes of operations were not observed is a consequence of the fact that the qubits on the IBM quantum processors have very similar resonant frequencies. In order for the engine to work, e.g., as a cooler, qubits with substantially different level spacings are necessary. Our results then, clearly indicate that in order to cool down qubits on a quantum processor, using a SWAP quantum heat engine, it is crucial to equip them with ``service'' qubits of larger level spacings, that can be used to extract energy from cold ``computational'' qubits. 
Such thermodynamic mechanism could be employed, e.g., to improve the purity of the computational qubits preparation, by sacrificing the purity of the service qubits.

\acknowledgments{We acknowledge the use of IBM Quantum services for this work \cite{IBMQ_ref}. The views expressed are those of the authors, and do not reflect the official policy or position of IBM or the IBM Quantum team. In this paper we used \emph{ibmq\_belem}, which is one of the IBM Quantum Falcon r4 Processors. 

Andrea Solfanelli and Alessandro Santini acknowledge that their research has been conducted within the framework of the Trieste Institute for Theoretical Quantum Technologies (TQT).
}

%

\end{document}